\documentstyle[honnef,pslatex]{article}  
\frompage{000} \topage{000}                                              
\newcommand{\beq}{\begin{equation}}
\newcommand{\eeq}[1]{\label{#1} \end{equation}}
\newcommand{\beqar}{\begin{eqnarray}}
\newcommand{\eeqar}[1]{\label{#1} \end{eqnarray}}
\newcommand{\dlt}{\bigtriangleup}

\def\rightmark{Phase transition in high energy 
heavy ion collisions ...} 
\def\leftmark{L. Csernai et al.}
 
\title{Phase Transitions in High Energy 
Heavy Ion Collisions within Fluid Dynamics} 
\authors{
{
L.P. Csernai$^{1,2}\footnote{Talk presented at the Symposium on Fundamental 
in Elementary Matter, Bad Honnef, Germany, September 25-29, 2000
}$, Cs. Anderlik$^{1}$ and V. Magas$^{1}$
}\\[2.812mm]
{\normalsize
\hspace*{-8pt}$^1$ Section for Theoretical and Computational Physics, 
Department of Physics\\University of
Bergen, Allegaten 55, 5007 Bergen, Norway\\
\hspace*{-8pt}$^2$ KFKI Research Institute for Particle and Nuclear
Physics\\P.O.Box 49, 1525 Budapest, Hungary\\[0.2ex]
}}
 
\abstract{Recent advances in Fluid Dynamical 
modeling of heavy ion collisions
are presented, with particular attention to  mesoscopic systems,
QGP formation in the pre FD regime and QGP hadronization
coinciding with the final freeze-out.\\
}
 
\begin{document}
 
\maketitle
\vspace*{24pt}

\section{Fluid Dynamics, and Local Equilibrium}
\label{sec1}

Fluid dynamics (FD) is probably the most frequently used model to
describe heavy ion collisions. It assumes local equilibrium, i.e. the
existence of an Equation of State (EoS), relatively short range
interactions and conservation of energy and momentum as well as
of conserved charge(s). Thus, it is a
widely usable model.

It can be derived most simply from the Boltzmann transport equation
where, the conservation laws, assuming local (approximate) kinetic
equilibrium yield the equations of (viscous) fluid dynamics.  If the
local momentum distribution deviates strongly from local kinetic
equilibrium the fluid dynamical approach (at least the one fluid one) is
not applicable.  This makes many believe that fluid dynamics has less
applicability than transport models, like molecular dynamics models.

One tends to forget other assumptions in transport models, i.e.  dilute
systems with binary collisions only, and consequently binary collisions.
These, constraints limit the applicability of transport models, for
example phase transitions (which include strong correlations, dense
systems, and not only binary collisions) can hardly be described
correctly with transport models.

Classical FD models incorporate phase transitions in a trivial way,
as the EoS, is given for a phase transition. Even more so, the FD
approach can describe systems out of phase equilibrium, supplemented
with a dynamical equation describing the dynamics of the phase 
transition, as local kinetic equilibrium in each phase
is sufficient to apply FD.\cite{CK92D,CK92L}

Phase transition dynamics is an involved question, even in {\it macroscopic
systems}. First of all, phase transitions can be different. This
expression may include, slow burning or deflagration, detonation,
condensation, evaporation, and many other forms of transition. The basic
conditions of all these transitions have, nevertheless, some
similarities.  These arise from the basic conservation laws and from the
requirement of equilibrium.

If we consider macroscopic stationary systems, asymptotically both the
initial and final states are in local mechanical (pressure), thermal and
chemical (or phase) equilibrium. The spatial extent and time-span of
the transitional region depends, on the other hand, from many features of
the transition.

Most explicit dynamical calculations are performed for the homogeneous
nucleation geometry as this is usually the mechanism which starts
the transition and which is the slowest of
all. 

In a dynamical situation the approach using the EoS including a first
order phase transition is identical both in situations involving
compression or expansion.  If the compression is supersonic, shockwaves
or detonation waves are formed, where the final new phase is immediately
formed. The phase transition speed influences only the width of the
shock front, but for slow dynamics and rapid phase transition the
shock front width is primarily determined by the transport coefficients,
viscosity and heat conductivity, and not by the phase transition speed.

If phase transitions occur in {\it small finite systems}
other dynamical features and configurations may occur as the dominant
form of a phase transition.

In small systems the system usually expands into the vacuum, and
freezes out, thus the final state is out of thermal and mechanical
equilibrium. Fluid dynamics cannot be applied at and after freeze-out
and even in a short period before freeze-out, when the assumption of
local thermal and mechanical equilibrium are not fully satisfied.
Connections of freeze-out and hadronization are discussed in section
\ref{sec3}.

In small finite systems among the configurations of possible
instabilities we cannot neglect the ones associated by the outer surface
of the system, which is of negligible importance for large macroscopic
systems but may be dominant for small systems.  

\section{Macroscopic Phase Transition Dynamics}
\label{sec2}

\subsection{Slow dynamics - rapid phase transition}

When we calculate the speed of the phase transition proper, we have
varying constraints and conditions. The speed of phase transition comes
into question only if the dynamics of the evolution otherwise is so fast
that it competes or exceeds the phase transition speed.

If the external speed is slow, we have sufficient time to have a
quasi-static process and reestablish phase equilibrium at every stage of
the dynamics.  This also means that all other equilibration processes are
also completed as these require less time and less interaction than
phase transition dynamics.

Thus, in the case of a "slow" external dynamics and rapid phase
equilibration the matter is in complete equilibrium, including phase
equilibrium, and the EoS of the matter having a first order phase
transition is given by the Maxwell construction: we have a fully
developed mixed phase, and the phase abundances are given by the fluid
dynamical evolution.  No extra information on dynamical processes is
needed.

\subsection{Deviation from phase equilibrium}

Even in moderately fast dynamical situations we have small deviations
from the ideal and complete phase equilibrium (Maxwell construction).
This deviation leads to some delay in the creation of the new phase
leading to supercooling or superheating, and extra entropy production.

For heavy ion reactions the first attempt to explicitly evaluate the 
phase transition speed of the homogeneous nucleation process is described
in refs. \cite{CK92D,CK92L}. The homogeneous nucleation mechanism
describes correctly the initial phases of the phase transition, where
the abundance of the newly created phase is still small, and when the 
phase transition process is the slowest.

Here a couple of remarks are necessary. To form bubbles or phases of the
new phase of supercritical size one needs to establish several requirements.
Pressure and temperature balance should be reestablished among the 
phases and this requires to establish the phase boundary and transfer
the needed energy and momentum to the new phase. We cannot relax the 
requirement of pressure and temperature equilibrium if both before and
after the formation of the new phase we assume local equilibrium and
so fluid dynamical evolution.

\subsection{Initial state}

As mentioned above the phase transition speed does not come into play
for really large, slow stationary systems. If the dynamics is supersonic
shock or detonation fronts will have the same initial and final
parameters, and only the shock front profile may be affected by the 
phase transition dynamics.

This is the typical type of approach also in heavy ion reactions up to
a few hundred $A\cdot MeV$ colliding energy.

\section{Phase Transition Dynamics in Small Systems}
\label{sec3}

\subsection{Final hadronization at freeze-out from QGP}

In principle it is possible that our system freezes out before
kinetic and/or phase equilibrium is established. Then, we end
up in a system of two or more phases, where none of them is
equilibrated kinetically, so we have no partial pressures and
temperatures.

The case of freeze-out directly from QGP is very special, because
hadronization must always be completed by the end of freeze-out, as QGP
or free quarks and gluons never reach the detectors. So, we must have
completed phase transition to a single phase even if no thermal or
mechanical equilibrium is established by the freeze-out.

Recently final freeze-out and hadronization was discussed in a series of
publications \cite{CLM97,ALM99,ACG99,MAC99hip,MAC99pl,MAC99npa}
improving essentially the previous Cooper-Frye freeze-out description
\cite{CF75} triggered by ref. \cite{Bugaev96} which suggested a 
solution to a long
standing problem in freeze-out description, but did not offer a complete
solution.

\begin{figure}[htb]
                 \insertplot{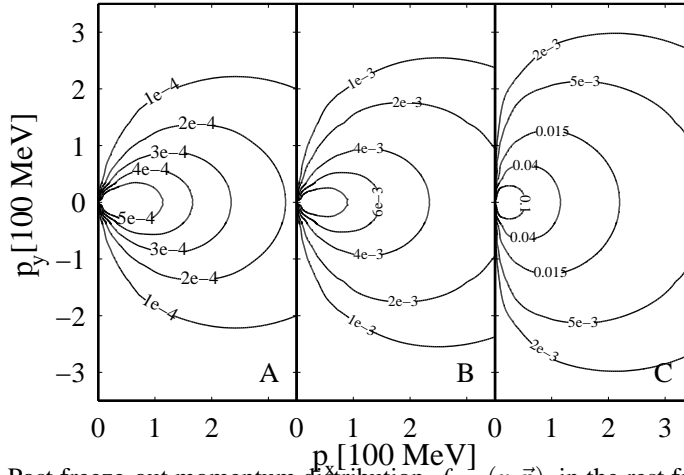}
\vspace*{-1cm}
\caption[]{Post freeze-out momentum distribution, $f_{free}(x,\vec{p})$, 
in the rest frame of the
freeze-out front obtained in 
\cite{MAC99hip}. Gradual kinetic freeze-out along $x$
 axis is described in a kinetic model.
 A, B and C correspond
to  freeze-out distances $0.2 \lambda,\ 3 \lambda,\  100 \lambda$  
respectively and
$u_{RFG}^\mu|_{x=0}=(1,0.5,0,0)$, where RFG - rest frame of gas, $\lambda$ is
scaling parameter of the order of mean free path.
The numbers in the contours are in arbitrary units. The distribution is
asymmetric and elongated in the freeze-out 
direction, $x$. This may lead to a large-$p_t$ enhancement, compared 
to the usual J\"uttner assumption used in most previous
calculations as a freeze-out distribution.
}
\label{freeze}
\end{figure}

In this series of works it is proven that in space-like ($d\sigma^\mu
d\sigma_\mu = -1$) freeze-out the post freeze-out local momentum
distribution cannot be a thermal distribution (because $p_\mu
d\sigma^\mu \ge 0$ should be satisfied),  and the earlier suggested 
cut-J\"uttner distribution is physically inadequate. A simple kinetic
model calculation provided an alternative non-equilibrium momentum
distribution, $f(p,e,n,u^\mu,d\sigma^\mu)$ (see Fig. \ref{freeze}).
 Using this distribution it
was shown that the freeze-out problem can be solved avoiding all
previously mentioned problems.

Most importantly the conservation laws must be exactly satisfied:
\beq
\left[ T^{\mu\nu} d\sigma_\mu \right] = 0 , \quad  
\left[ N^{\mu} d\sigma_\mu \right] = 0 ,\quad  
\left[ S^{\mu} d\sigma_\mu \right] \ge 0 ,\quad 
\eeq{eqa3}
where the square bracket stands for the difference of the post and
pre freeze-out values. Neither these conditions nor the adequate
post freeze-out non-equilibrium distributions were used in the
earlier Cooper-Frye freeze-out models. If these conditions are satisfied 
at the freeze-out surface and the post freeze-out distribution, 
$f(p,e,n,u^\mu,d\sigma^\mu)$ is properly chosen the so called Cooper-Frye
{\em formula} can still be used.

The post freeze-out non-equilibrium distribution should nevertheless 
be evaluated in an adequate nonequilibrium dynamical model. The simple 
kinetic models used in refs. \cite{ALM99,ACG99,MAC99hip,MAC99pl,MAC99npa}
 can be considered as a
first step only, which does not address important further details as
post freeze-out flavor abundances, rapid and sequential freeze-out
mechanisms, etc. A realistic and detailed freeze-out model requires a
consistent dynamical model which generates such a distribution.

Present string and parton cascade models if they can handle both
the pre and post freeze-out phases realistically (!)  may be used for this
purpose with proper care.  First attempts of coupling hadronic string
models to the end of FD are performed already. \cite{BD00} However,
these models do not handle the phase transition dynamics, so satisfying
the last condition, the requirement of entropy increase, is nontrivial in
this model.

Note that sudden or rapid hadronization from QGP with entropy increase
usually possible only in most model calculations if the plasma is
sufficiently supercooled.

\subsection{Phase transition mechanisms}

The mechanisms of phase transitions have a large
variety and consequently their dynamical features are also different.
Experiments on strangeness and on two particle correlations suggest for
some years by now, that hadronization and freeze-out is a rapid process,
as the final observed system size is small and strange particle
abundances are large. This contradicts to homogeneous processes in
thermal and approximate phase equilibrium which were studied in the
beginning of the 90s.

In ref. \cite{CC94} it was demonstrated that rapid QGP hadronization
with small volume increase is possible even if we require energy- and
momentum conservation and non-decreasing entropy. However, in ref.
\cite{CC94} no explanation was presented which would enable such a
transition, which poses a problem as the earlier proposed homogeneous
nucleation is not fast enough to support such a rapid transition.
Alternative homogeneous processes (e.g. spinodal decomposition) were not
studied in detail but as far as near thermal equilibrium is maintained
it is difficult to imagine that these can support a qualitatively faster
process.

The need to find other processes which can support rapid hadronization
was obvious by the mid 90s. The first attempt was to fully relax the
requirement of thermalization, even the existence of temperature, and
find a non-thermal, field theoretical, mechanism for the hadronization
\cite{CM95}. This connection
gave then renewed activity in the study of fluctuations in field
theories and in study of DCC. However, most of these studies were still
considering a homogeneous transition.

\section{Phase Transition in the Initial State} 

At highly ultra-relativistic energies, when QGP is formed, the
pre collision initial state is far out of any thermal or mechanical
equilibrium, and equilibrium can only be established in the QGP phase
where the number of degrees of freedom are sufficiently large and the
interaction frequency is also large so that equilibration is expected to
be established in a tenth of a $fm/c$. 

Nevertheless, the preceding dynamics cannot be treated by any thermal
or fluid dynamical model, and we need QCD based effective models
which originate from the observations gained from particle and
heavy ion physics experiments at these energies.

Just like connecting the hydrodynamic stage and freeze-out to each
other on a 3 dimensional hypersurface a detailed description of an
energetic heavy ion reaction requires a {\it Multi Module Model},
 where the
different stages of the reaction are each described with suitable
theoretical approaches. It is important that these Modules are coupled
to each other correctly: on the interface, which is a 3 dimensional
hyper-surface in space-time with normal $d\sigma^\mu$, all conservation
laws should be satisfied and entropy
should not decrease, eq.(\ref{eqa3}). These matching conditions were
worked out and studied for the matching at FO in detail in refs.
\cite{CLM97,ALM99,ACG99,MAC99hip,MAC99pl,MAC99npa}.

The initial stages are the most problematic. Frequently two or three fluid
models are used to remedy the difficulties, and to model the process of
QGP formation and thermalization. \cite{A78,C82,bsd00} Here, the problem
is transferred to the determination of drag-, friction- and transfer-
terms among the fluid components, and a new problem is introduced with
the (unjustified) use of EoS in each component in a nonequilibrated
situations, where EoS does not exist. Strictly speaking this approach
can only be justified for mixtures of noninteracting ideal gas
components.  Similarly, the use of transport theoretical approaches
assuming dilute gases with binary interactions is questionable, as due
to the extreme Lorentz contraction in the C.M. frame enormous particle
and energy densities with the immediate formation of perturbative
vacuum should be handled. Even in most parton cascade models these
initial stages of the dynamics are just assumed in form of some initial
condition, with little justification behind.

All string models had to introduce new, energetic objects: string ropes
\cite{bnk84,S95}, quark clusters \cite{WA96}, fused strings \cite{ABP93},
in order to describe the abundant formation of massive particles like
strange antibaryons.  Based on this, we describe the initial moments of
the reaction in the framework of classical (or coherent) Yang-Mills
theory, following ref. \cite{GC86} assuming larger field strength
(string tension) than in ordinary hadron-hadron collisions.  In addition
we now satisfy all conservation laws exactly, while in ref.  \cite{GC86}
infinite projectile energy was assumed, and so, overall energy and
momentum conservation was irrelevant. 

\subsection{Coherent Yang-Mills model}

Our basic idea is to generalize the model developed in \cite{GC86}, for
collisions of two heavy ions and improve it by strictly satisfying
conservation laws.  First of all, we would create a grid in $[x,y]$
plane ($z$ -- is the beam axes, $[z,x]$ -- is reaction plane).  We will
describe the nucleus-nucleus collision in terms of steak-by-streak
collisions, corresponding to the same transverse coordinates, $\{x_i,
y_j\}$.  We assume that baryon recoil for both target and projectile
arise from the acceleration of partons in an effective field
$F^{\mu\nu}$, produced in the interaction.  Of course, the physical
picture behind this model should be based on chromoelectric flux tube or
string models, but for our purpose we consider $F^{\mu\nu}$ as an
effective abelian field. Phenomenological parameters describing this
field must be fixed from comparison with experimental data.

Let describe the streak-streak collision.
\beq
\partial_\mu \sum_i T_i^{\mu\nu}=\sum_i F_i^{\nu\mu} n_{i \mu} \ ,
\eeq{eq1}
\beq
\partial_\mu \sum_i n_i^\mu = 0 \ , \quad i=1,2\  ,
\eeq{eq2}
$n_i^\mu$ is the baryon current of $i$th nucleus (we are working in the
Center of Rapidity Frame (CRF), which is the same for all streaks.  The
concept of using target and projectile reference frames has no advantage
any more).  We will use the parameterization:
\beq
n_i^\mu=\rho_i u_i^\mu \ ,
\quad
u_i^\mu=(\cosh y_i,\ \sinh y_i)  \ .
\eeq{eq3}
$T^{\mu\nu}$ 
is a energy-momentum flux tensor. It
consists of five parts, corresponding to both
nuclei and free field energy 
$T^{\mu\nu}_{F,i}$ 
(also divided into two parts) and one
defining the QGP perturbative vacuum.\\
\beq
T^{\mu\nu}=\sum_i T_i^{\mu\nu}+T^{\mu\nu}_{pert}=
\sum_i\left[ e_i\left(\left(1+c_0^2\right)u_i^\mu u_i^\nu
- c_0^2g^{\mu\nu}\right)
+T_{F,i}^{\mu\nu}\right]+B g^{\mu\nu}\ ,
\quad i=1,2 \ ,
\eeq{eq4}
where $B$ -- is the bag constant, the equation of state is $P_i=c_0^2
e_i$, where $e_i$ and $P_i$ are energy density and pressure of QGP.
In complete analogy to electro-magnetic field
\beq
F_i^{\mu\nu}=\partial^\nu A_i^{\mu}-\partial^\mu A_i^{\nu}=\left(
\begin{array}{cc}
0 & -\sigma_i \\
\sigma_i & 0
\end{array}\right) \ \ , {\rm where} \ \ 
\sigma_i=\partial^3 A_i^{0}-\partial^0 A_i^{3}\ ,
\eeq{eq7}
\beq
T_{F,i \mu\nu}=-g_{\mu\nu}{\cal L}_{F,i}+\sum_\beta \frac{{\cal L}_{F,i}}
{\partial \left(\partial^\mu A_i^{\beta}\right)}\partial_\nu A_i^{\beta}
\ \ , {\rm where}\ \ 
{\cal L}_{F,i} = - \frac{1}{4}F_{i \mu\nu}F_i^{\mu\nu}\ .
\eeq{eq9}
The string tensions, $\sigma_i$, have the same absolute value $\sigma$ 
and opposite sign (in complete analogy to the usual string with two ends
moving in opposite directions), and $\sigma_i$ will be constant in the
space-time region after string creation and before string decay.

We received analytic solutions of the above equations using light cone
variables \cite{GC86}, $x^{\pm} = t \pm z$. Following \cite{MCS00,incond} we
assume that target variables, $e_1,\ y_1,\ \rho_1,\ A^\mu_1$ are
functions of $x^-$ only and projectile variables of $x^+$ only. At the
time of first touch of two streaks, $t=0$, there is no string tension
built up yet.  We assume that strings are created, i.e. the sting
tension achieves the value $\sigma$ at time $t=t_0$, corresponding to
complete penetration of streaks through each other. 

\subsubsection{Conservation laws --- string rope creation}

In light cone variables the baryon current conservation, eq. (\ref{eq2})
may be rewritten as
\beq
\partial_-n_1^-+\partial_+n_2^+=0 \ .
\eeq{eq19}
So, we have a sum of two terms, depending on different 
independent variables,
and the solution can be found in the following way:
$
\partial_-n_1^-=a, \
\partial_+n_2^+=-a \ ,
$
with
$
n_1^-=a x^- + (n_1)_0, \ \ 
n_2^+=-a x^+ + (n_2)_0 \ .
$
Since both $n_1^-$ and $n_2^+$ are positive (and also more or less symmetric) 
we can conclude that for
our case $a=0$.
Finally:
$
n_1^-=\rho_1 e^{-y_1}=\rho_0 e^{y_0} \ , \quad
n_2^+=\rho_2 e^{y_2}=\rho_0 e^{y_0} \ ,
$
and
$
\rho_1=\rho_0 e^{y_0+y_1}\ , \quad  \rho_2=\rho_0 e^{y_0-y_2}  \ .
$

As mentioned before, after string creation,
i.e. $t>t_0$, and before string decay we choose the string 
tensions in the
form:
$
\sigma_2=-\sigma_1=\sigma>0\ .
$ 
With this choice and the Lorentz gauge condition we take the 
vector potentials in the following form \cite{MCS00,incond}:
\beq
A_1^+ = 0, \quad A_1^- = -2 \sigma x^- \ ; \quad
A_2^+ = -2 \sigma x^+ , \quad A_2^- = 0 \ ,
\eeq{eq24}
where we used the parameterization:
\beq
\sigma=A\left(\frac{\varepsilon_0}{m}\right)^2\rho_0
\sqrt{l_1l_2} \ ,
\eeq{eq44}
where $l_1$ and $l_2$ are the initial streak lengths. 
We are working in the system, 
where $\ \hbar=c=1$. 
The typical values of dimensionless parameter 
$A$ are around $0.045-0.055$. The above parameterization is arbitrary 
in the sense that the requirements of the right dimension and 
grid size independence do not completely fix it. Above 
parameterization has been checked to work 
in the energy range 
$\varepsilon_0=10-3000 \ GeV$ per nucleon. Notice, that
there is only one free parameter in parameterization (\ref{eq44}).
The typical values of $\sigma$ are
$5-12\ GeV/fm$ for $\varepsilon_0=100 \ GeV$
per nucleon.
These values are consistent with the energy density
in non-hadronized strings, or "latent energy density" which is on the
average 8 GeV/fm$^3$.\cite{ASC91pl,ASC91prl,ACS92}

As eq. (\ref{eq1}) has a source term we do not know what the really
conserved quantities are. 
Using the solution for $n_1^-$ and $n_2^+$, we can define new energy-momentum
tensor $\tilde{T}^{\mu\nu}$, such that
\beq
\partial_\mu \tilde{T}^{\mu\nu}=0  , \quad
\tilde{T}^{\mu\nu}=\sum_i \tilde{T_i}^{\mu\nu}+T^{\mu\nu}_{pert}=
\sum_i \left(T_i^{\mu\nu} - A_i^\nu n_i^\mu +
g^{\mu\nu} A_i^\alpha n_{i \alpha} \right)+B g^{\mu\nu}
\eeq{eq27}

Now the new conserved quantities are
\beq
Q_0=\int \tilde{T}^{00} dV =\dlt x\dlt y \sum_i \int_{l_i} \tilde{T_i}^{00} dz
\ ,
\quad
Q_3=\int \tilde{T}^{03} dV =\dlt x\dlt y \sum_i \int_{l_i} \tilde{T_i}^{03} dz
\ ,
\eeq{eq29}
where $\dlt x\dlt y$ is the cross section of the streaks.

Then the trajectories of nucleons (or cell elements)
for both nuclei, the energy density and baryon density distributions
can be obtained analytically and are given in \cite{MCS00,incond}.

\subsubsection{Recreation of the matter}

As we may see from the trajectories
nucleons (or cell domains) will keep going in the initial direction up
to the time $t=t_{i,turn}$, then they will turn and go backwards until
the two streaks again penetrate through each other and new oscillation
will start. Such a motion is analogous to the "Yo-Yo" motion in the
string models.  Of course, it is difficult to believe that such a
process would really happen in heavy ion collisions, because of string
decays, string-string interactions, interaction between streaks and
other reasons, which are quite difficult to take into account.  To be
realistic we should stop the motion 
at some moment before the projectile and target cross again.

We assume that the final result of collisions of two streaks after
stopping the string's expansion and after its decay, is one streak of the 
length $l_f$ with
homogeneous energy density distribution, $e_f$, and baryon charge
distribution, $\rho_f$, moving like one object with rapidity $y_f$. We
assume that this is due to string-string interactions and string decays.
As it was mentioned above the typical values of the string tension,
$\sigma$, are of the order of $10\ GeV/fm$, and these may be treated as
several parallel strings. The string-string interaction will produce a
kind of "string rope" between our two streaks, which is responsible for
final energy density and baryon charge homogeneous distributions.  Now
it is worth to mention that decay of our "string rope" does not allow
charges to remain at the ends of the final streak, as it would be if we
assume full transparency.

The homogeneous distributions are the simplest assumptions, which may be
modified based on experimental data. Its advantage is a simple
expression for $e_f,\ \rho_f,\ y_f$.

The final energy density, baryon density and rapidity, $e_f, \ \rho_f$ and
$y_f$, should be determined from conservation laws. Unfortunately, the 
assumptions we made above oversimplify the real situations and do not 
allow us to satisfy exactly all the conservation laws. The 
reason for this is well known and has been discussed in the 
Refs. \cite{ALM99,ACG99,MAC99hip,MAC99pl,MAC99npa}:  
two possible definitions of the flow, Eckart's and Landau's definitions,
are not always identical. 
The exact conservation of 
the energy and momentum gives for the final rapidity:
\beq
\cosh^2 y_{f,L} = \frac{(M^2(1+c_0^2)+2c_0^2u_0^2)+
\sqrt{(M^2(1+c_0^2)+2c_0^2u_0^2)^2+4c_0^4u_0^2(M^2-u_0^2)}}{2(1+c_0^2)
(M^2-u_0^2)} \ ,
\eeq{eq66}
where we neglected $B \dlt l_f$ next to $Q_0/\dlt x\dlt y$ 
and introduced the 
notation
$M=(l_2+l_1)/(l_2-l_1)$, $u_0=\tanh y_0$ is the initial velocity.
The exact conservation of the baryon four-current gives:  
\beq
\tanh y_{f,E} = \frac{u_0}{M} \ , \rightarrow \  
\cosh^2 y_{f,E} = \frac{M^2}{M^2-u_0^2} \ .
\eeq{eq66b}
It is interesting that if we put $c_0^2=0$ the eq. (\ref{eq66}) becomes 
identical to eq. (\ref{eq66b}). For more details see \cite{incond}.

We follow the Refs. \cite{MCS00,incond}, 
where the $y_{f}=y_{f,L}$ has been chosen. 
In this case the expressions for the $e_f$ and $\rho_f$ are:
\beq
e_f=\frac{{Q_0\over\dlt x \dlt y}}{\left((1+c_0^2)\cosh^2 y_f 
- c_0^2\right) l_f} \ ,
\eeq{eq68}
\beq
\rho_f=\frac{\rho_0(l_1+l_2)}{l_f \cosh y_f} \ .
\eeq{eq69}

So, the streaks  move until they reach 
the rapidity $y_i=y_f$. 
Later the final streak starts to move like one object with rapidity $y_f$.

\subsection{Initial conditions for hydrodynamical calculations}
 
\begin{figure}[htb]
                 \insertplot{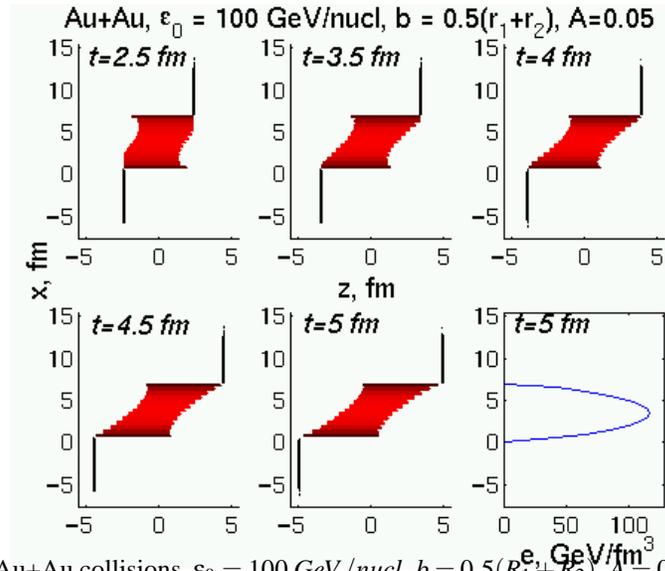}
\vspace*{-1cm}
\caption[]{The Au+Au collisions, $\varepsilon_0=100\ GeV/nucl$, 
$b=0.5(R_1+R_2)$, $A=0.05$ (parameter $A$ introduced in (\ref{eq44})),
 $y=0$ (ZX plane through 
the centers of nuclei). We would like to notice that 
final shape of QGP volume is a tilted disk $\approx 45^0$, 
and the direction of the fastest expansion will deviate from both 
the beam axis and the usual transverse flow direction, and might be a 
reason for the third flow component, as argued in \cite{CR99}.
}
\label{ev11}
\end{figure}

We are interested in the shape of QGP formed, when string 
expansions stop and
their matter is locally equilibrated.  This will be the initial state
for further hydrodynamical calculations. We may see in Figs. \ref{ev11},
that QGP forms a tilted disk for $b\not =0$. So, the direction of
fastest expansion, the same as largest pressure gradient, will be in the
reaction plane, but will deviate from both the beam axis and the usual
transverse flow direction.  So, the new flow component, called
"antiflow" or "third flow component", may appear in addition to the
usual transverse flow component in the reaction plane.  With increasing
beam energy the usual transverse flow is getting weaker, while this new
flow component is strengthened. The mutual effect of the usual directed
transverse flow and this new "antiflow" or "third flow component" leads
to an enhanced emission in the reaction plane.  This was actually
observed and studies earlier. One should also mention that both the 
standard transverse flow and new "antiflow" contribute to the
 "elliptic flow".

\section{Critical Fluctuations}

Fluid Dynamics inherently describes the average behavior out the
thermodynamical ensemble characterizing a system. Event by event
fluctuations can nevertheless be included in FD, and one can generate
an ensemble of FD events. For mesoscopic systems the fluctuations
are not negligible, and near the critical point these can even dominate
the dynamics.\cite{LCJ97,CJK97,CLL00} Critical fluctuations may
signal the vicinity of the critical point in a phase transition,
thus may serve as a QGP signal. 

One may use the Dissipation-Fluctuation Theorem \cite{CJK97} to generate
fluctuating Langevin forces on physical ground. Note that although this
problem for the case of Navier-Stokes equation was addressed already by
Landau in 1957, a practically usable approach was only worked out 
recently.

In heavy ion reactions the latent heat of the phase transition released
at the final hadronization may feed fluctuations, among other possibilities
also critical fluctuations in FD. The first methodological steps
were made in this direction, but a realistic and experimentally verifiable
prediction of observable fluctuations is still some way ahead.

\section{Conclusions}

We recalled arguments for a rapid freeze-out in heavy ion collisions,
which coincides with the hadronization of QGP, and presented a considerably 
improved method for the idealized description of the freeze-out.

Based on earlier Coherent Yang-Mills field theoretical models, and
introducing effective string tension 
parameters based on Monte-Carlo string cascade and
parton cascade model results, a simple model is introduced to
describe the pre fluid dynamical stages of heavy ion collisions at the
highest SPS energies and above.  The model predicts limited transparency
for massive heavy ions,
 as a consequence of collective effects related to QGP formation.
These collective effects in central and semi central collisions lead to
 much less transparency than earlier estimates. The resulting
initial locally equilibrated state of matter in semi central collisions
takes a rather unusual form, which can be then identified by the
asymmetry of the caused collective flow.  Our prediction is that this
special initial state may be the cause of the recently predicted
"antiflow" or "third flow component". 


\vfill\eject
\end{document}